\documentclass[aip,jcp,reprint]{revtex4-1}
\renewcommand{\vec}[1]{\mbox{\boldmath$#1$}}

\newcommand{\me}{\mathrm{e}}

\newcommand{\dif}{\mathrm{d}}

\usepackage{graphicx}

\usepackage{amsmath}

\usepackage{amsfonts}

\usepackage{dcolumn}

\usepackage{bm}

\begin{document}

\title{Multi-scale theory in the molecular simulation of electrolyte
solutions}
\author{W. Zhang}\email{wzhang4@tulane.edu}
\author{X. You}\email{xyou@tulane.edu}
\author{L. R. Pratt}\email{lpratt@tulane.edu}
\affiliation{Department of Chemical and Biomolecular Engineering, Tulane
University, New Orleans, LA 70118}

             
\begin{abstract} 
This paper organizes McMillan-Mayer theory, the potential distribution
approach, and quasi-chemical theory to provide theory for the
thermodynamic effects associated with longer spatial scales involving
longer time scales, thus helping to define a role for AIMD simulation
directly on the time and space scales typical of those demanding
methods.   The theory treats composition fluctuations which would be
accessed by larger-scale calculations, and also longer-ranged
interactions that are of special interest for electrolyte solutions. The
quasi-chemical organization breaks-up governing free energies  into
physically distinct contributions: \emph{packing}, \emph{outer-shell},
and \emph{chemical} contributions.  Here we study specifically the
\emph{outer-shell} contributions that express electrolyte screening. For
that purpose we adopt a primitive model suggested by observation of
ion-pairing  in tetra-ethylammonium tetra-fluoroborate dissolved in
propylene carbonate. Gaussian statistical models are shown to be
effective physical models for \emph{outer-shell} contributions, and they
are conclusive for the free energies within the quasi-chemical
formulation.  With the present data-set the gaussian physical
approximation obtains more accurate mean activity coefficients than does
the Bennett direct evaluation of that free energy. \end{abstract}

\maketitle

\section{Introduction}

This paper develops statistical mechanical theory with the goal of treating
electrolyte solutions at chemical resolution. Our context is current
research on electrochemical double-layer capacitors (EDLCs) based on
nanotube forests.\cite{Yang:2010hd} The requirement of chemical
resolution means that electronic structure must be an integral part of
the theory consistent with the natural interest in chemical features of
EDLCs.

\emph{Ab initio} molecular dynamics (AIMD), though not statistical
mechanical theory, is available to simulate electrolyte solutions.
Compared to classic molecular simulations with empirical model
force-fields, AIMD calculations are severely limited in  space and time
scales, by more than an order-of-magnitude in each.   Consequently,
application of AIMD is not feasible for EDLCs at scales that are
experimentally interesting.   This calls for further theory to embed
AIMD methods in studies of EDCLs.

Change-of-scale consequences are a primitive goal of fundamental
statistical mechanical theory.  That basic perspective is explicit in
the classic phase transition literature,\cite{Ma} and it has long been
relevant to the theory of electrolyte solutions
specifically.\cite{hfrie77,hfrie81}   This paper organizes several basic
results of the statistical mechanics of solutions to treat electrolyte
solutions where space and time scales will otherwise prohibit direct
AIMD calculations.  Our results here suggest a role for AIMD somewhat
analogous to sub-grid modeling in computational fluid mechanics.
Nevertheless, the goals of the statistical thermodynamics of complex
solutions are distinct, and we do not propose transfer of results here
between those fields.

\begin{widetext}
\begin{center}
\begin{figure}[h]
\includegraphics[width=6.5in]{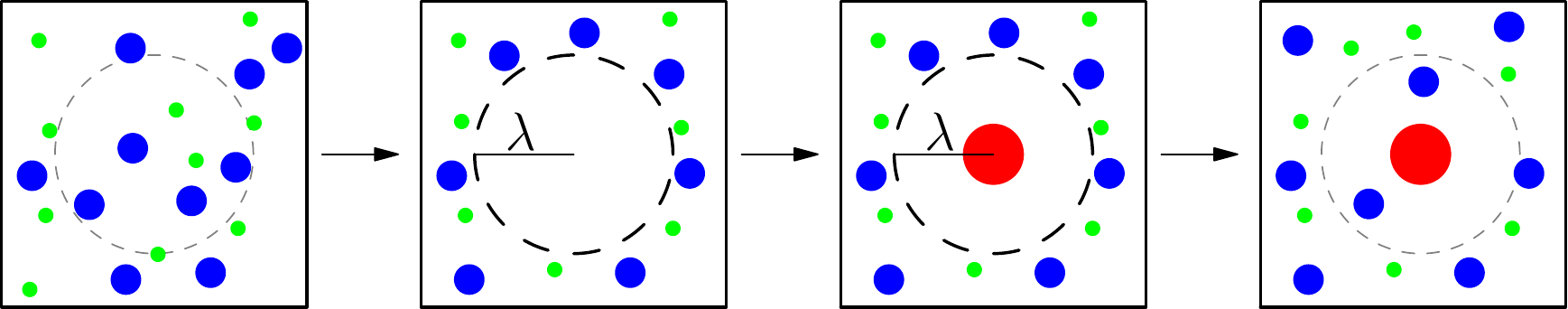}
\caption{Evaluation of the excess chemical potential of a distinguished
ion (red disk), patterned according to QCT. The blue and green disks are other ions
in the system, and the solvent is in the background. The stepwise
contributions are  ``packing,'' ``outer shell,'' and ``chemical''
contributions, from left to right. See the text and
Eq.~\eqref{eq:BIGQCT} for further discussion.}
\label{fig:QCT}
\end{figure}
\end{center}
\end{widetext}

Our development here utilizes several theoretical results that are
``\ldots both difficult and strongly established \ldots'' \cite{hfrie77}
We put burdensome technical results in appendices, and in this
introduction give a fuller discussion of the line of reasoning.

The initial step in our development is the McMillan-Mayer (MM)
theory\cite{wmcmi45,thill60,hfrie77} \emph{integrating out} of solvent
degrees of freedom.   MM theory is a pinnacle of coarse-graining for the
statistical mechanics of solutions, all primitive models rest on it, and
it achieves a vast conceptual simplification the theory of electrolyte
solutions. No sacrifice of molecular realism is implied by MM theory.
But cataloging the multi-body potentials required for a literal MM
application is prohibitive.\cite{sadel76} Therefore, use of MM theory 
to construct a specific primitive model for a system of experimental
interest has been limited.\cite{pkusa88,curse91} Indeed, the MM theory 
is not generally suitable for specific molecular-scale implementation.

To address this we exploit \emph{quasi-chemical theory} (QCT) which is
formally complete in its modern
expression.\cite{Asthagiri:2010tj,sabo2013} QCT evaluates solvation free
energies by breaking them into contributions with clear physical
meanings.  One contribution is a \emph{packing} contribution.   This can
be simple in the anticipated applications because the solvent is not
involved specifically, and the ion concentrations are not prohibitively
high. A second contribution --- the \emph{outer shell} contribution ---
treats ion-ion interactions at long-range and it is expected on physical
grounds that the necessary MM interactions should be simple then.  That
outer-shell contribution is studied below.

The final contribution --- the \emph{chemical} contribution --- treats
ion-ion inner-shell neighbors.  Smaller spatial scales must be directly
confronted and it is here that the sub-grid AIMD activity comes into
play. Fig.~\ref{fig:QCT} shows a now-standard picture of this
organization of the statistical thermodynamical problem.

This discussion suggests that van~der~Waals theory is a subset of the
present QCT approach.  This is advantageous because van~der~Waals theory
is the basis of the theory of liquids viewed
broadly.\cite{Widom:1967tz,Chandler:1983vr}  While paying an unavoidable
price of  significant computational effort, QCT goes beyond
van~der~Waals theory in  several ways. For example here, where
ion-pairing is an essential part of the physical picture, associative
phenomena are  treated fully. Furthermore, our QCT implementation would
routinely treat outer-shell interactions through gaussian order rather
than the mean-field approach of classic van~der~Waals theories. This is
essential in the present applications in order to capture the physical
effect of Debye screening of ion correlations.

It is an interesting physical point that the identification of
\emph{packing} and \emph{chemical} contributions here is a consequence
of a choice of conditioning event, in the present development the
emptiness of the inner-shell.   This has the advantages that the theory
is a close relative of van der Waals theory, and that the outer-shell
contribution should be particularly simple to evaluate.  But other
choices of conditioning event are possible too.\cite{Rogers2012}  For
example, the conditioning event might be the event that the occupancy of
the inner-shell is the value most probably observed. This has the
intuitive attraction of  being  close to simple observations.  But it
presents the challenge that the evaluation of the partition function for
that case might be more difficult.   In what follows, our primary
emphasis is to characterize the computational effort to evaluate the
partition function associated with \emph{outer-shell} contribution that
arises with the original suggestion for the conditioning event.

The plan of this paper is as follows:   Sec.~\ref{sec:BTR} records
several theoretical specifics that are required for our argument.  The
Appendix A gives an accessible derivation of the MM theory results used
here; Appendices B and C present technical features of the Potential
Distribution Theory and Quasi-Chemical Theory, respectively, required in
the main text. Sec.~\ref{sec:Demo} gives a
demonstration of the  results obtained for a primitive electrolyte
solution model that was designed to correspond to the
tetra-ethylammonium tetra-fluoroborate in propylene carbonate
(TEABF$_4$/PC) where ion-pairing can be important.\cite{Zhu:2011wn}

\bigskip
\section{Basic Theory Required}\label{sec:BTR}
\subsection{McMillan-Mayer theorem}\label{sec:MM}
The osmotic pressure, $\pi$, is evaluated as the partition function
\begin{eqnarray}
\me^{\beta \pi V} =
\sum_{\vec{n}_\mathrm{A}\ge0}
\mathcal{Z}\left(\vec{n}_{\mathrm{A}};\vec{z}_\mathrm{S}\right)
\left(\frac{\vec{z}_\mathrm{A}{}^{\vec{n}_\mathrm{A}}}{\vec{n}_\mathrm{A}!}\right)~,
\label{eq:MMpf}
\end{eqnarray}
involving only the solute species A.  Here $V$ is the volume,  
$k_\mathrm{B}T = \beta^{-1}$ the temperature, and the activity of the 
solvent (species S) is denoted by 
$z_{\mathrm{S}}=\me^{\beta\mu_{\mathrm{S}}}$.    Eq.~\eqref{eq:MMpf} 
involves
\begin{multline}
\mathcal{Z}\left(\vec{n}_{\mathrm{A}};\vec{z}_\mathrm{S}\right) = 
\left\lbrack\lim_{\vec{z}_{\mathrm{A}}\rightarrow 0}\left(\frac{\rho_\mathrm{A}}{z_\mathrm{A}}\right)^{\vec{n}_{\mathrm{A}}}\right\rbrack \\
\times \int_V \dif 1_\mathrm{A}
\ldots \int_V \dif n_\mathrm{A}\me^{ - \beta W\left( 1_\mathrm{A} \ldots
n_\mathrm{A} \right)}~,
\label{eq:MMpfFINAL}
\end{multline}
with $\rho_\mathrm{A}$ being the density of solutes, and with the potentials-of-average-force given by
\begin{eqnarray}
W\left( 1_\mathrm{A} \ldots
n_\mathrm{A} \right) =
-\frac{1}{\beta}\ln g\left( 1_\mathrm{A} \ldots
n_\mathrm{A};\vec{z}_\mathrm{S},\vec{z}_{\mathrm{A}}=0 \right)~,
\end{eqnarray}
which depends on the the activity of the solvent.
The result Eq.~\eqref{eq:MMpfFINAL} is compact, thermodynamically 
explicit, and general;  see the Appendix for an accessible derivation 
and fuller discussion.

\subsection{Potential distribution theorem (PDT)}\label{sec:PDT}

The solute chemical potential may be expressed as 
\begin{multline}
\beta \mu_\mathrm{A} = \ln \rho_\mathrm{A} \Lambda_\mathrm{A}{}^3/ q_\mathrm{A}^{\left(\mathrm{int}\right)}
+ \beta \mu_{\mathrm{A}}^{\left(\mathrm{ex}\right)}\left(z_\mathrm{A}=0\right) \\
- \ln\left\langle\left\langle
\me^{-\beta \Delta W^{(1)}_\mathrm{A}}
\right\rangle\right\rangle_0 ~.
\label{eq:MMfinal}
\end{multline}
The binding energy of a distinguished solute (A) molecule in the MM
system is
\begin{eqnarray}
\Delta W^{(1)}_\mathrm{A} = W\left(\vec{n}_A+1\right) - W\left(\vec{n}_A\right) - W\left(1\right)~.
\end{eqnarray}
The middle term of Eq.~\eqref{eq:MMfinal}, 
\begin{eqnarray}
\beta \mu_{\mathrm{A}}^{\left(\mathrm{ex}\right)}\left(z_\mathrm{A}=0\right) = - \ln \left\langle\left\langle
\me^{-\beta \Delta U^{(1)}_\mathrm{A}}\right\rangle
\right\rangle_0~,
\end{eqnarray}
is evaluated at infinite dilution of the solute. This evaluation is
typically highly non-trivial, but much has been written about
that\cite{Asthagiri:2010tj,sabo2013} and we will proceed to analyze that
right-most term of Eq.~\eqref{eq:MMfinal}.

\subsection{Quasi-chemical theory}\label{sec:QCT}

Thus we study
\begin{multline}
\beta \Delta\mu_{\mathrm{A}}^{\left(\mathrm{ex}\right)} = 
\beta \mu_{\mathrm{A}}^{\left(\mathrm{ex}\right)} - 
\beta \mu_{\mathrm{A}}^{\left(\mathrm{ex}\right)}\left(z_\mathrm{A}=0\right) \\
= - \ln\left\langle\left\langle
\me^{-\beta \Delta W^{(1)}_\mathrm{A}}
\right\rangle\right\rangle_0
 ~.
\label{eq:MMGC}\end{multline}
$\Delta\mu_{\mathrm{A}}^{\left(\mathrm{ex}\right)}$ is the contribution
to the chemical potential of species A in excess of the infinite
dilution result, due to inter-ionic interactions with the influence of
solvent fully considered.

A quasi-chemical development of Eq.~\eqref{eq:MMGC} starts by
characterizing neighborship. If the species considered are ions in
solution, then we need to characterize ion neighbors of each ion in
solution, distinguished in turn.  Pairing of oppositely charged ions has
been the subject of classic scientific history\cite{Zhu:2011wn} that can
inform the present discussion.  Pairing of tetra-fluoroborate
1-hexyl-3-methylimidazolium in pentanol has recently been studied both
experimentally and computationally.\cite{Zhu:2012eo}

Pairing of tetra-ethylammonium tetra-fluoroborate in propylene carbonate
is a helpful example.\cite{Zhu:2011wn}   In that case, pairing is simple
to observe for saturated solution conditions and formation of chains and
rings of ions is consistent with the molecular-scale observations.   We
might consider an indicator function $\chi_\mathrm{AB}$ with the
requirement that $\chi_{\mathrm{AB}}=1$ indicates \emph{no} B ions are
within an inner-shell stencil of a distinguished A ion.  The simplest
possibility, natural for compact molecular ions, is to identify a
central atom in A and in B ions, and to set  $\chi_{\mathrm{AB}}=1$ (but
zero otherwise), when those atoms of further apart than a designated
distance $\lambda_{\mathrm{AB}}$.

Even simpler, and satisfactory for the primitive model that follows
below, we might identify a central atom for ion A, then define a
spherical inner-shell by the radius $\lambda_{\mathrm{A}}$ with the
requirement that \emph{no} other ions be closer than that.   In fact, we
choose the same radius for cations and anions in the primitive model
studied below, $\lambda_{\mathrm{A}}=\lambda_{\mathrm{A}}=\lambda$.

Eq.~\eqref{eq:MMGC} was derived using the grand canonical ensemble. But
implementation with simulations in the grand canonical ensemble would be
painful.   Calculations with \emph{canonical} ensemble methods should be
satisfactory. In what follows we will develop Eq.~\eqref{eq:MMGC} from
the perspective on the canonical ensemble.  Appendix C discusses the
relevant  ensemble differences.

The canonical ensemble average of $\me^{\beta \Delta W^{(1)}_\mathrm{A}}\chi_{\mathrm{A}}$ gives
\begin{multline}
\left\langle
\me^{\beta \Delta W^{(1)}_\mathrm{A}}\chi_{\mathrm{A}}
\right\rangle
=\frac{\left\langle\left\langle
\me^{-\beta \Delta W^{(1)}_\mathrm{A}}\me^{\beta \Delta W^{(1)}_\mathrm{A}}\chi_{\mathrm{A}}
\right\rangle\right\rangle_0}{\left\langle\left\langle
\me^{-\beta \Delta W^{(1)}_\mathrm{A}}
\right\rangle\right\rangle_0} \\
=
\me^{\beta \Delta\mu_{\mathrm{A}}^{\left(\mathrm{ex}\right)}}\left\langle\left\langle
\chi_{\mathrm{A}}
\right\rangle\right\rangle_0~.
\label{eq:conditioning}
\end{multline}
In addition, for  indicator function $\chi_{\mathrm{A}}$, and some other quantity
$G$ we have 
\begin{eqnarray}
\left\langle
G\chi_{\mathrm{A}}
\right\rangle
=\left\langle
G |\chi_{\mathrm{A}} = 1
\right\rangle\left\langle
\chi_{\mathrm{A}}
\right\rangle~.
\end{eqnarray}
Collecting these relations yields
\begin{eqnarray}
\me^{\beta \Delta\mu_{\mathrm{A}}^{\left(\mathrm{ex}\right)}} = \left\langle
\me^{\beta \Delta W^{(1)}_\mathrm{A}} |\chi_{\mathrm{A}} = 1
\right\rangle \times
\frac{\left\langle
\chi_{\mathrm{A}}
\right\rangle}{\left\langle\left\langle
\chi_{\mathrm{A}}
\right\rangle\right\rangle_0}~.
\end{eqnarray}
After evaluating the logarithm and replacing the conditional ensemble
average with an integral over a conditional probability distribution, we get
\begin{multline}
\beta \Delta\mu_{\mathrm{A}}^{\left(\mathrm{ex}\right)} = -\ln\left\langle\left\langle
\chi_{\mathrm{A}}
\right\rangle\right\rangle_0 \\
 + \ln\int\me^{\beta \varepsilon} P_{\mathrm{A}}\left( \varepsilon | \chi_{\mathrm{A}} = 1 \right)d\varepsilon + \ln \left\langle
\chi_{\mathrm{A}}
\right\rangle~
\label{eq:BIGQCT}
\end{multline}
where
\begin{eqnarray}
P_{\mathrm{A}}\left( \varepsilon | \chi_{\mathrm{A}} = 1 \right) = \left\langle
\delta\left(\varepsilon - \Delta W^{(1)}_{\mathrm{A}}\right) | \chi_{\mathrm{A}} = 1 \right\rangle~.
\end{eqnarray}
These three terms correspond to the three processes in Fig.~\ref{fig:QCT}.

\section{Numerical demonstration}\label{sec:Demo}

\begin{table}
  \centering
    \begin{tabular}{| c | c | c | c|c|}
    \hline
    
	$c~\left(\mathrm{mol/dm^3}\right)$  & 
    $L$  (nm) & 
    $n_{\mathrm{ion-pairs}}$ &  
    $\kappa^{-1}$ (nm) &
    ${\beta q^2\kappa}/{2\epsilon}$ \\ \hline
    0.01 & 32.15 & 200 &  2.67 & 0.17\\ \hline
    0.05 & 18.80 & 200 &  1.19 & 0.39\\ \hline
    0.1 & 14.92 & 200 &  0.84 & 0.55\\ \hline
    0.2 & 11.84 & 200 &  0.60 & 0.77\\ \hline
    0.4 & 9.4 & 200 &   0.42 & 1.11\\ \hline
    0.5 & 9.4 & 250 &   0.37 & 1.26\\ \hline
    0.6 & 9.4 & 300 &  0.34  & 1.37\\ \hline
    0.8 & 9.4 & 400 &  0.30  &  1.55\\  \hline
    1.0 & 9.4 & 500 &  0.27  &  1.72\\ \hline
    2.0 & 7.4 & 500 &  0.19  & 2.44\\    
    \hline
    \end{tabular}
   \caption{Specifications for Monte Carlo simulation of a primitive
   model with dielectric constant, ion charges and sizes corresponding
   to the atomically detailed [TEA][BF$_4$]/PC case.\cite{Zhu:2011wn}  Specifically
   the model dielectric constant is $\epsilon = 60$, and $d_{++} =
   0.6668$~nm, $d_{--} = 0.6543$~nm, $d_{-+} = 0.45$~nm are distances of
   closest approach for the hard spherical ions.  These calculations
   utilized the Towhee\cite{Towhee} package adapted to the present
   system, conventional cubical periodic boundary conditions at $T$ =
   300K, and the indicated concentrations $c$.
   Each calculation was extended to 10$^6$ cycles after aging, each
   cycle comprising $2n_{\mathrm{ion-pairs}}$ attempted moves.  10,000 configurations
are saved and used for the following analyses. }
   \label{table:table2}
\end{table}

\begin{figure}
\center{\includegraphics[width=1.0\linewidth]{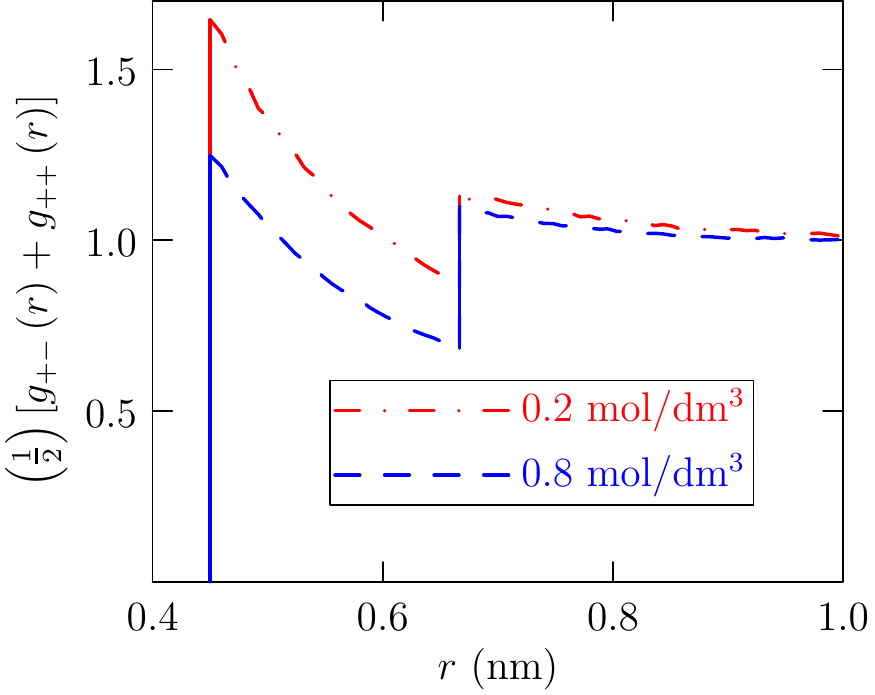}}
\caption{The radial distribution functions of $c$ = 0.2 mol/dm$^3$ 
and $c$ = 0.8 mol/dm$^3$ from cation to other ions. These two 
vertical lines identify the closest approach distances, which are 
0.45~nm and 0.6668~nm in this case.
}
\label{fig:fullcon6}
\end{figure}
\begin{figure}
\includegraphics[width=3.5in]{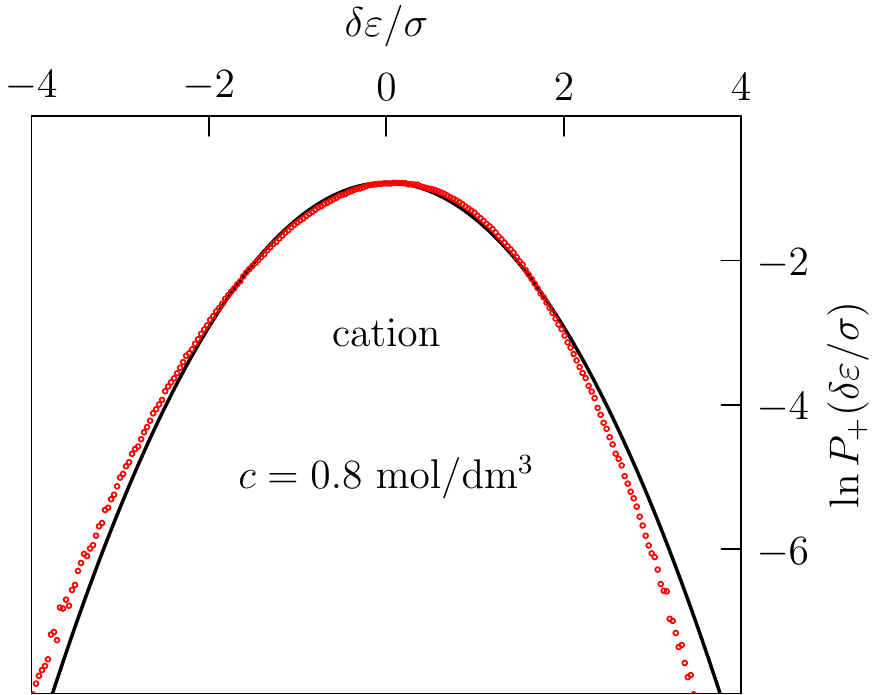}
\includegraphics[width=3.5in]{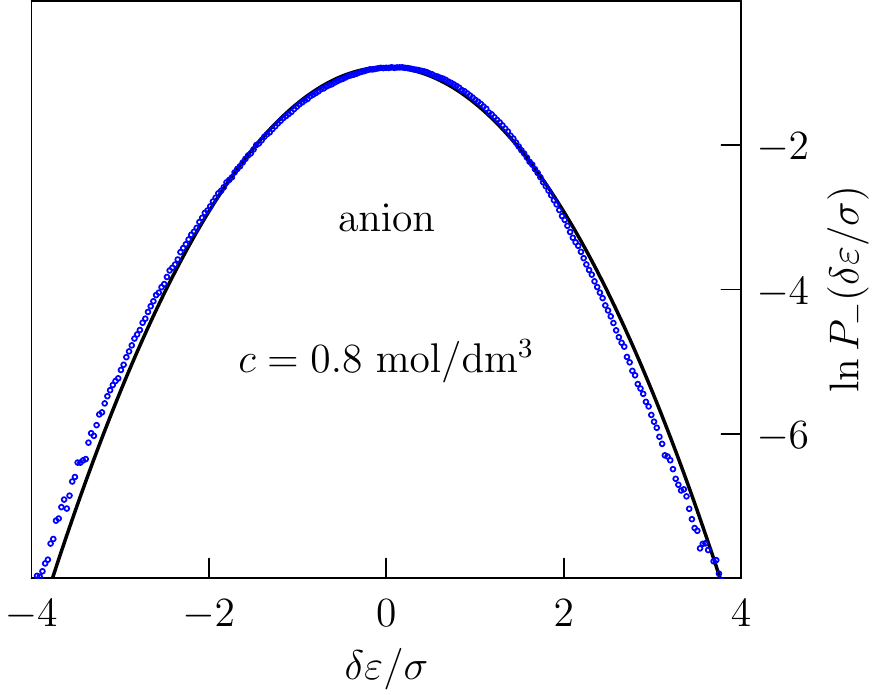}
\caption{Observed distributions of ion binding energies, shifted and
scaled into standard normal form, for $c$ = 0.8~mol/dm$^3$. The
parabolae (solid black lines) here are standard normal comparisons.
}
\label{fig:grConditional2}
\end{figure}

\begin{figure}
\includegraphics[width=3.5in]{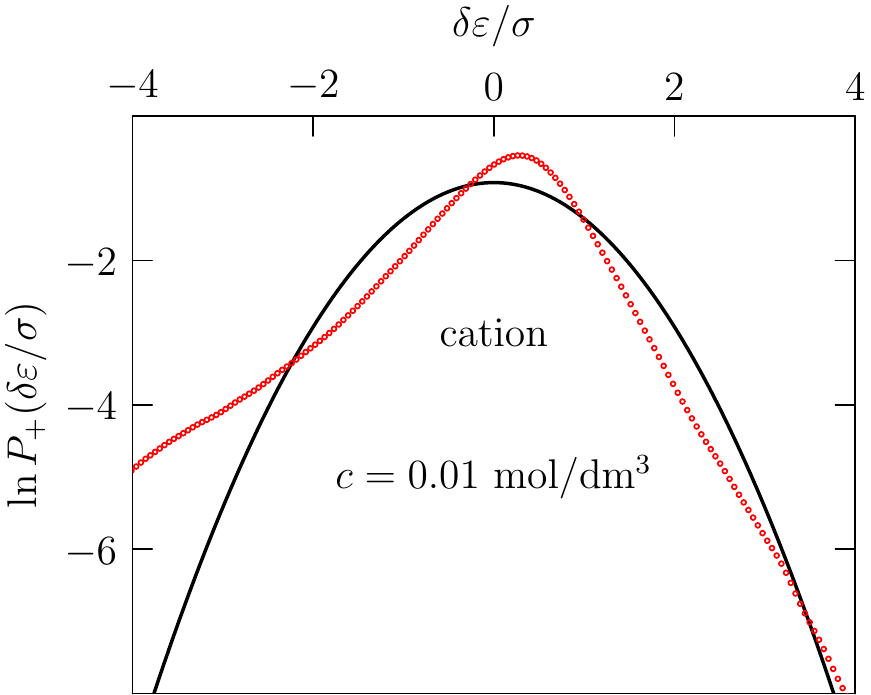}
\includegraphics[width=3.5in]{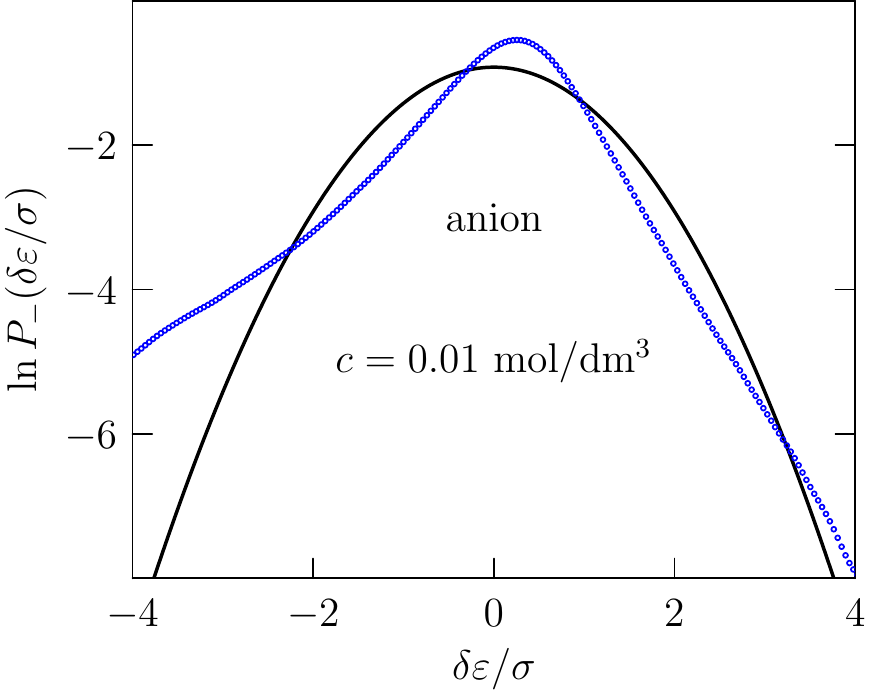}
\caption{Observed distributions of ion binding energies, shifted and 
scaled into standard normal form, for $c$ = 0.01~mol/dm$^3$. The 
parabolae (solid black lines) here are standard normal comparisons.
}
\label{fig:grConditional1}
\end{figure}

\begin{figure}
\includegraphics[width=3.5in]{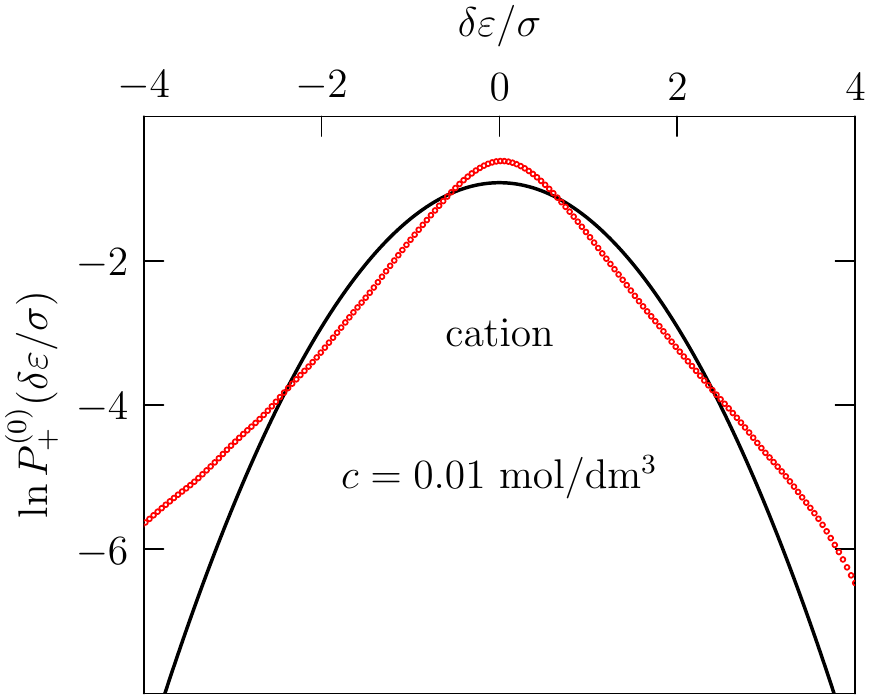}
\includegraphics[width=3.5in]{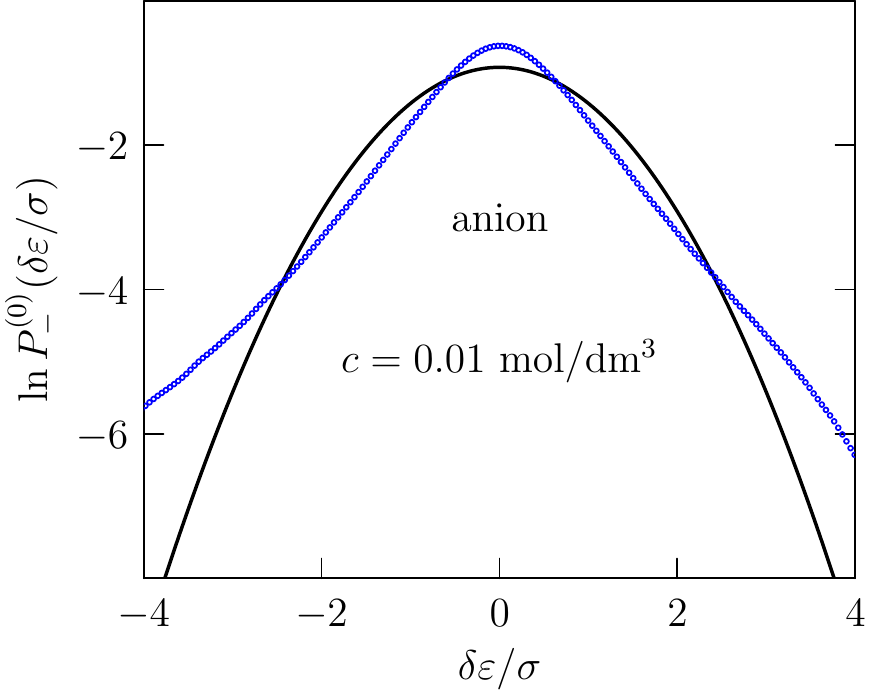}
\caption{Observed distributions of uncoupled ion binding energies,
shifted and scaled into standard normal form, for 
$c$ = 0.01~mol/dm$^3$. The parabolae (solid black lines) here are 
standard normal comparisons.
}
\label{fig:grConditional3}
\end{figure}

\begin{figure}
\includegraphics[width=3.5in]{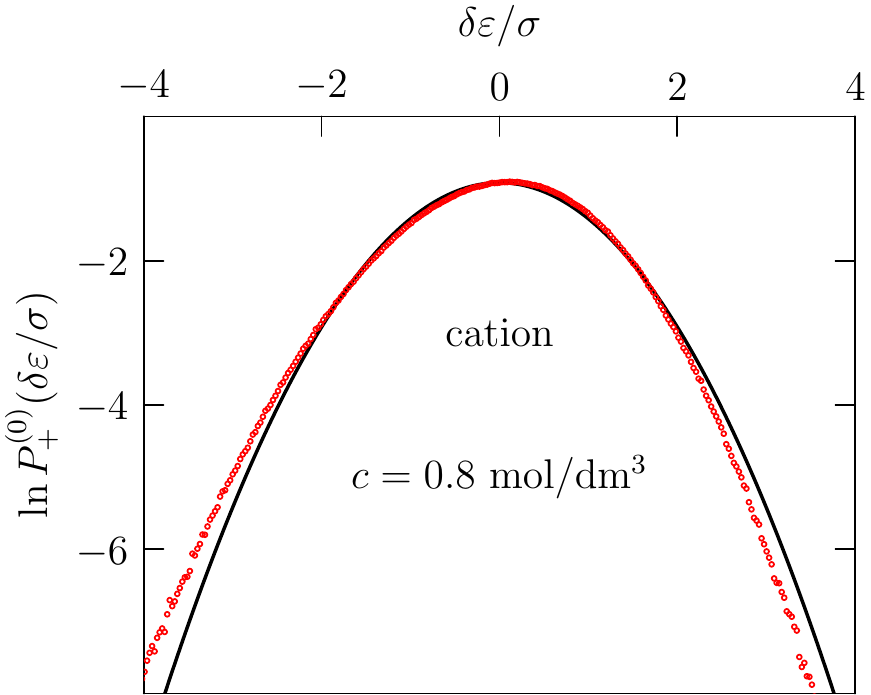}
\includegraphics[width=3.5in]{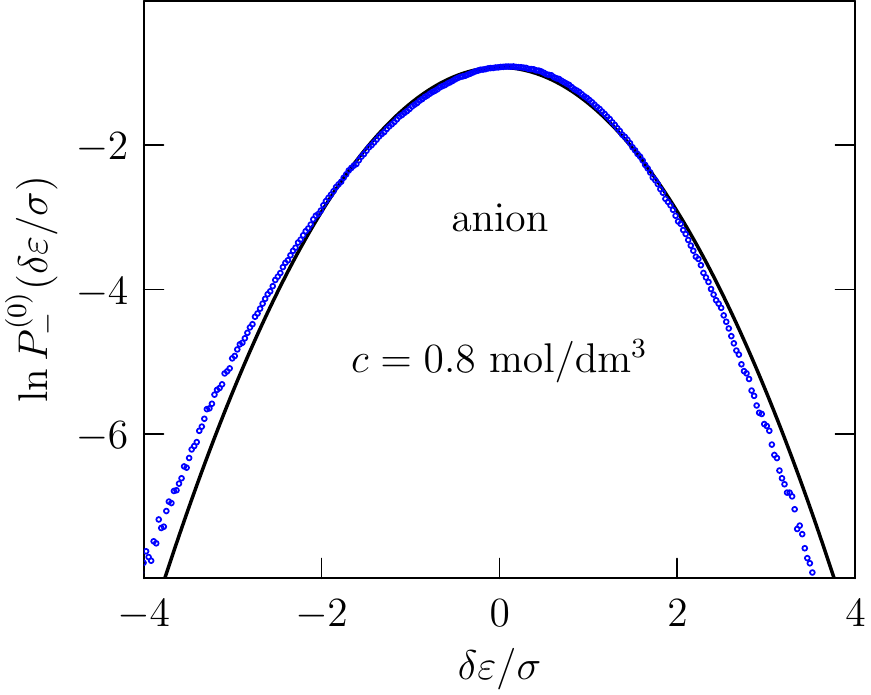}
\caption{Observed distributions of uncoupled ion binding energies,
shifted and scaled into standard normal form, for 
$c$ = 0.8~mol/dm$^3$. The parabolae (solid black lines) here are 
standard normal comparisons.
}
\label{fig:grConditional4}
\end{figure}

\begin{figure}
\includegraphics[width=3.5in]{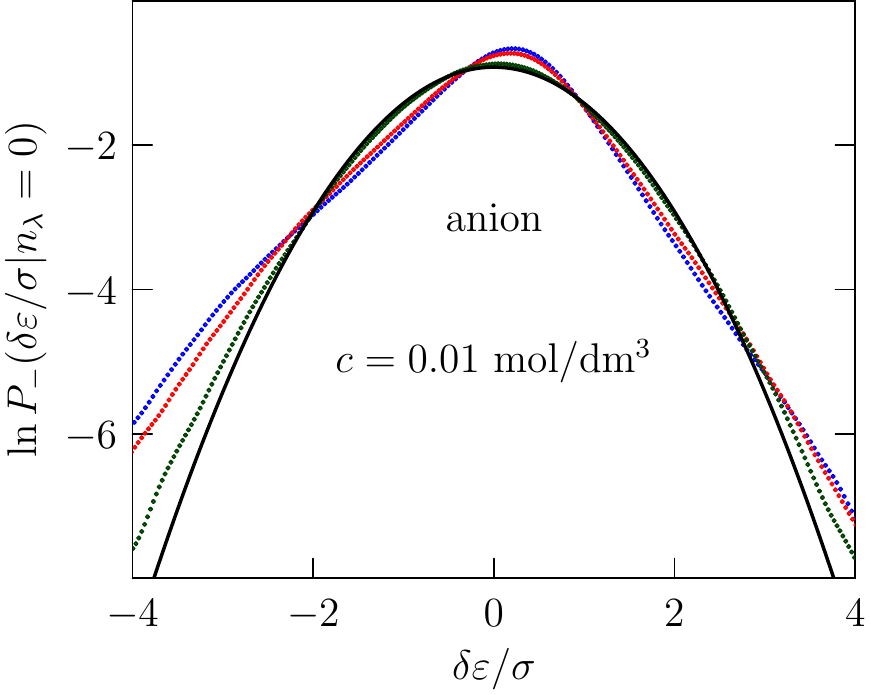}
\caption{Probability density functions for the outer-shell binding
energy for the anions in the simulation of Table~\ref{table:table2},
for the lowest concentration there, with $\lambda$ = 0.7, 0.9, and
2.0~nm (blue, red, and darkgreen respectively), compared the standard
normal (solid black curve).  This demonstates how increasing 
$\lambda$ enforces better Gaussian behavior for this distribution.
}
\label{fig:fitXAnion10mM}
\end{figure}

We have tested  how Eq.~\eqref{eq:BIGQCT} works numerically on the basis
of a primitive electrolyte solution model that was designed to
correspond to the TEABF$_4$/PC. Table~\ref{table:table2} describes the
model further and indicates the thermodynamic states studied by Monte
Carlo calculations.  The radial distribution functions
(Fig.~\ref{fig:fullcon6}) show why this primitive model, with
non-additive hard-sphere interactions, was identified to study
ion-pairing.

In this case, the packing and chemistry contributions can be directly
calculated by trial insertions (for \emph{packing} and
$\left\langle\left\langle \chi_{\mathrm{A}} \right\rangle\right\rangle_0
$), and observation of the closest neighbor molecule distance
distribution (for \emph{chemistry} and $\left\langle \chi_{\mathrm{A}}
\right\rangle$). That leaves the \emph{outer-shell} contribution which
is our particular interest here.

For the general theory (Eq.~\eqref{eq:BIGQCT} and Fig.~\ref{fig:QCT}),
the distinguished ion will be separated by a substantial distance from
all other ions.  We assume that the required binding energy can be
approximated as a superposition of the pair potential-of-mean-force at
long-range between the distinguished ion and all ion neighbors, which we
take to be the classic macroscopic result $q_i q_j /4 \pi \epsilon r$
for a separation of $r$.   That superposition is just the electrostatic
inter-ionic potential energy of interaction for the primitive model
considered here.

\subsection{Binding energy distributions}
We evaluate binding energies for the primitive model by standard Ewald
calculation for configurations extracted from the Monte Carlo
simulations, $\varepsilon = \Delta W_{\mathrm{A}}^{(1)}$. We examine the
distributions of binding energies for the ions present in the
simulation, $P(\varepsilon )$, and also binding energies for permissible
trial placements of additional anions or cations, $P^{(0)}(\varepsilon
)$.  In order to have a basis for comparison, we do these calculations
first \emph{without} the conditioning prescribed by QCT, \emph{i.e.},
all anions or cations without regard to their neighborship status.
Distributions of those binding energies (Figs.~\ref{fig:grConditional2},
\ref{fig:grConditional1}, \ref{fig:grConditional3}, and
\ref{fig:grConditional4})  are striking. At the higher concentration
shown (Fig.~\ref{fig:grConditional2}), the distributions are reasonably
normal as expected. At the lower concentration shown
(Fig.~\ref{fig:grConditional1}), the distributions are non-gaussian. 
The design of the model to reflect ion-pairing is evident in the
enhanced weight at substantially negative binding energies.

The normal presentation (as in Figs.~\ref{fig:grConditional2},
\ref{fig:grConditional1}, and \ref{fig:grConditional3}, shift-scaled and
compared to standard normal) helps to judge the width of these
distributions. An alternatively presentation
(Figs.~\ref{fig:OverlapAnion10mM} and \ref{fig:OverlapAnion800mM})
compares these binding energy ranges to the thermal energy $kT$ and
gives additional insight. The free energy prediction from the coupled
distributions (Figs.~\ref{fig:grConditional2}, \ref{fig:grConditional1})
depends sensitively on the high-$\varepsilon$ (right) wing of these
graphs and hardly at all on the low-$\varepsilon$ (left) wing. Even
though the low-concentration distribution
(Fig.~\ref{fig:grConditional1}) is strikingly non-gaussian, the
right-wing extends, very roughly, to the same width as the natural
gaussian.   In contrast the uncoupled $P^{(0)}(\varepsilon )$
(Fig.~\ref{fig:grConditional3}) is qualitatively non-gaussian in both
high-$\varepsilon$ and low-$\varepsilon$ wings.

The  uncoupled $P^{(0)}(\varepsilon )$ (Figs.~\ref{fig:grConditional3})
at low concentration also distinctly abnormal,  in both wings; at
high-concentration they (Fig.~\ref{fig:grConditional4}) more nearly
gaussian.

\subsection{QCT conditioned binding energy distributions} We next
consider the conditioned distributions that arise with the QCT approach
(Eq.~\eqref{eq:BIGQCT} and Fig.~\ref{fig:QCT}).  We take the inner-shell
to be a sphere of radius $\lambda$ centered on the ions. Typical results
for $P_-({\delta\varepsilon}/{\sigma}\vert n_\lambda = 0)$ for the
interesting low concentration case (Fig.~\ref{fig:fitXAnion10mM}) shows
how the conditioning affects this distribution, with increasing
$\lambda$ driving the distribution toward normal behavior.

\subsection{Free energies and gaussian approximations}
The goal of our QCT development is to break the free energy into parts
associated first with simple observations, and finally with a partition
function calculation (the \emph{outer-shell} contribution) that can be
well  approximated by a gaussian model with simply observed
parameters. In that case the \emph{outer-shell} contribution of
Eq.~\eqref{eq:BIGQCT} would be  
\begin{multline}
 \ln\int\me^{\beta \varepsilon} P_{\mathrm{A}}\left( \varepsilon | \chi_{\mathrm{A}} = 1 \right)d\varepsilon \\
 \approx  \beta \langle  \varepsilon  | \chi_{\mathrm{A}} = 1 \rangle 
+  \beta^2 \langle \delta \varepsilon ^2 | \chi_{\mathrm{A}} = 1 \rangle/2 ~.
\label{eq:gapprox}
\end{multline}
To test these ideas, we evaluate the free energies directly, with and 
without the QCT conditioning, and also compare the results of the 
gaussian approximation  Eq.~\eqref{eq:gapprox}.

The direct evaluation of the free energies follows Bennett's
method,\cite{BENNETTCH:Effefe} and searches for the value  $\Delta
\mu^{\mathrm{(ex)}}$ that solves
\begin{eqnarray}
\left\langle
\frac{1 }{1+\mathrm{e}^{-\beta\left(\varepsilon -  \Delta \mu_{\mathrm{A}}^{\mathrm{(ex)}}\right )}} 
\right\rangle
=
\left\langle\frac{1}{1+\mathrm{e}^{\beta\left(\varepsilon -  \Delta \mu_{\mathrm{A}}^{\mathrm{(ex)}} \right )}}
\right\rangle_0~
\label{eq:Bmethod}
\end{eqnarray}
for each species considered.  The average on the left is
estimated with the sample associated with $P_{\mathrm{A}}(\varepsilon )$
whereas the average on the right of uses the binding energies leading to
$P_{\mathrm{A}}^{(0)}(\varepsilon )$ associated with permissible 
trial placements.

\subsubsection{No conditioning}\label{sec:nonQCT}
In this case, we estimate a non-electrostatic contribution directly 
by trial  insertions, then electrostatic contribution on the basis of 
distributions  such as Figs.~\ref{fig:grConditional1} and 
\ref{fig:grConditional3}. The mean activity coefficients 
(Fig.~\ref{fig:meangammaV5}) obtained with the gaussian 
approximation and the Bennett evaluation are qualitatively 
similar but quantitatively different from each other.

\subsubsection{Pointwise Bennett comparison}

\begin{figure}
\includegraphics[width=3.5in]{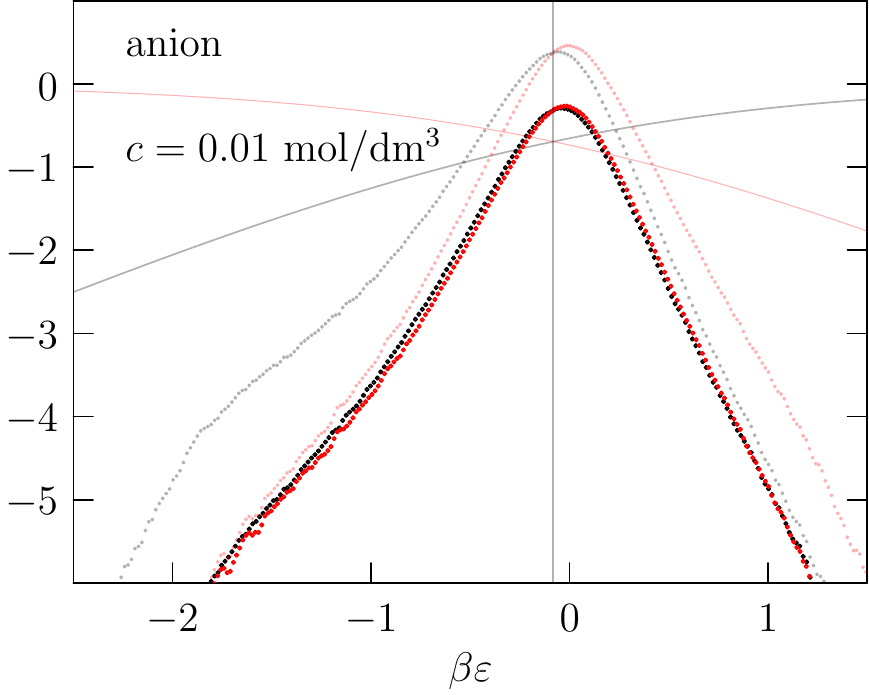}
\caption{ For the low concentration case of Table~\ref{table:table2}.
The bold circles are the logarithms of the functions left and right
of Eq.~\eqref{eq:BPs}, black and red, respectively. The pale solid
curves are the corresponding weight factors, \emph{e.g.} $-\ln\left\lbrack 1+
\me^{\beta\left(\varepsilon -\Delta \mu_\alpha^{\mathrm{(ex)}}
\right)}\right\rbrack$ for the red curve. The pale symbols are the plotted
logarithms of the observed probability densities. The vertical line is
the inferred value of the $\beta\Delta\mu_\alpha^{\mathrm{(ex)}}$.  Note the
Eq.~\eqref{eq:BPs} is accurately satisfied.}
\label{fig:OverlapAnion10mM}
\end{figure}

\begin{figure}
\includegraphics[width=3.5in]{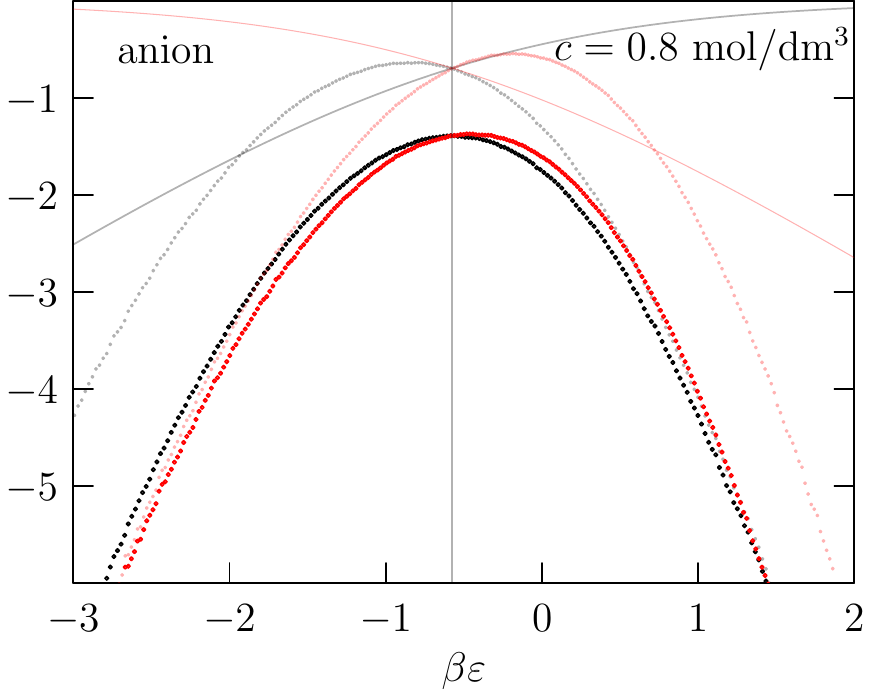}
\caption{For the $c$=0.8~mol/dm$^3$ case of Table~\ref{table:table2}.
The bold circles are the logarithms of the functions left and right
of Eq.~\eqref{eq:BPs}, black and red, respectively. The pale solid 
curves are the corresponding weight factors, \emph{e.g.} $-\ln\left\lbrack 1+
\me^{-\beta\left(\varepsilon -\Delta \mu_\alpha^{\mathrm{(ex)}}
\right)}\right\rbrack$ for the black curve. The pale symbols are the plotted
logarithms of the observed probability densities. The vertical line is
the inferred value of the $\beta\Delta\mu^{\mathrm{(ex)}}$. Note the
Eq.~\eqref{eq:BPs} is only roughly satisfied in the low-$\varepsilon$
wing.
}
\label{fig:OverlapAnion800mM}
\end{figure}

\begin{figure}
\includegraphics[width=3.5in]{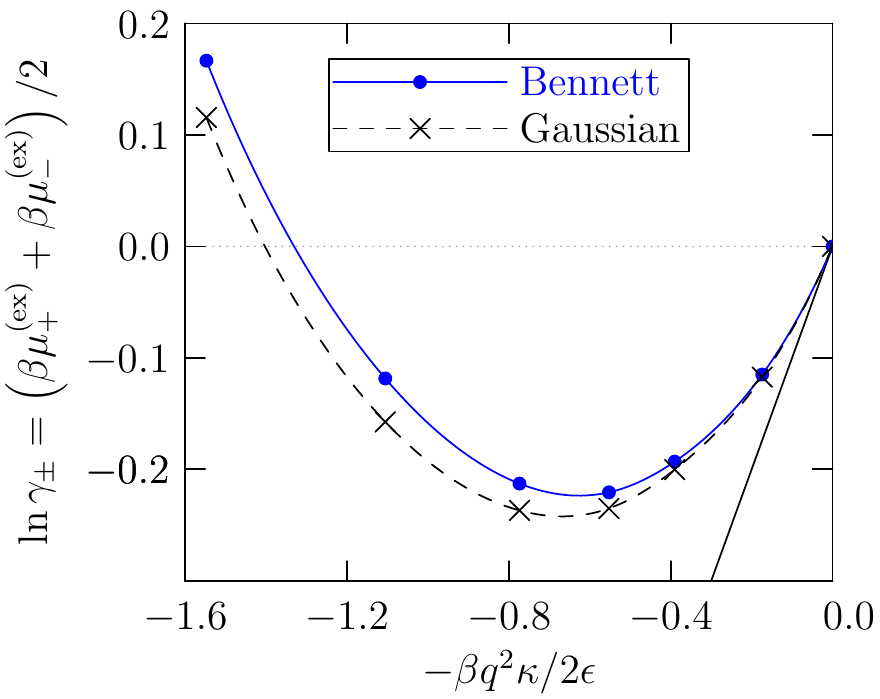}
\caption{No conditioning, non-QCT, as discussed in
Sec.~\ref{sec:nonQCT}. The solid black line is the Debye-H\"uckel
limiting law.
}
\label{fig:meangammaV5}
\end{figure}

A more specific statement of the Bennett approach is
\begin{eqnarray}
\frac{P\left(\varepsilon\right) }{1+\me^{-\beta\left(\varepsilon -  \Delta \mu^{\mathrm{(ex)}}\right )}} 
=
\frac{P^{\mathrm{(0)}}\left(\varepsilon\right) }{1+\mathrm{e}^{\beta\left(\varepsilon -  \Delta \mu^{\mathrm{(ex)}} \right )}}~.
\label{eq:BPs}
\end{eqnarray}
This relies on the basic relation
\begin{eqnarray}
P\left(\varepsilon\right)  = \me^{-\beta\left(\varepsilon -  \Delta \mu^{\mathrm{(ex)}}\right )}P^{\mathrm{(0)}}\left(\varepsilon\right)~,
\end{eqnarray}
and the elementary identity
\begin{eqnarray}
\me^{-\beta\left(\varepsilon -  \Delta \mu^{\mathrm{(ex)}}\right )}
=\frac{1+\me^{-\beta\left(\varepsilon -  \Delta \mu^{\mathrm{(ex)}}\right )} 
}{
1+\me^{\beta\left(\varepsilon -  \Delta \mu^{\mathrm{(ex)}} \right )}}~.
\end{eqnarray}
Eq.~\eqref{eq:BPs} assembles  information from
$P\left(\varepsilon\right)$ and $P^{(0)}\left(\varepsilon\right)$, and
thus illuminates the behavior of the important wings, high-$\varepsilon$
for $P\left(\varepsilon\right)$ and low-$\varepsilon$ for
$P^{(0)}\left(\varepsilon\right)$. Typical results
(Figs.~\ref{fig:OverlapAnion10mM} and \ref{fig:OverlapAnion800mM}) show
reasonable match between the left and right side of Eq.~\eqref{eq:BPs}.
It is helpful to note that at the higher concentration shown
(Fig.~\ref{fig:OverlapAnion800mM}), the match in the low-probability
wings is not perfect.  A reasonable guess is that this is due to
inaccuracy of $P^{(0)}\left(\varepsilon\right)$ at low-$\varepsilon$ and
that this is the reason behind the puzzling discrepancy between the
Bennett result and the gaussian model seen in
Figs.~\ref{fig:meangammaV5} and \ref{fig:meangammaV5lambda}.

\subsubsection{QCT conditioning}

\begin{figure}
\center{\includegraphics[width=1.0\linewidth]{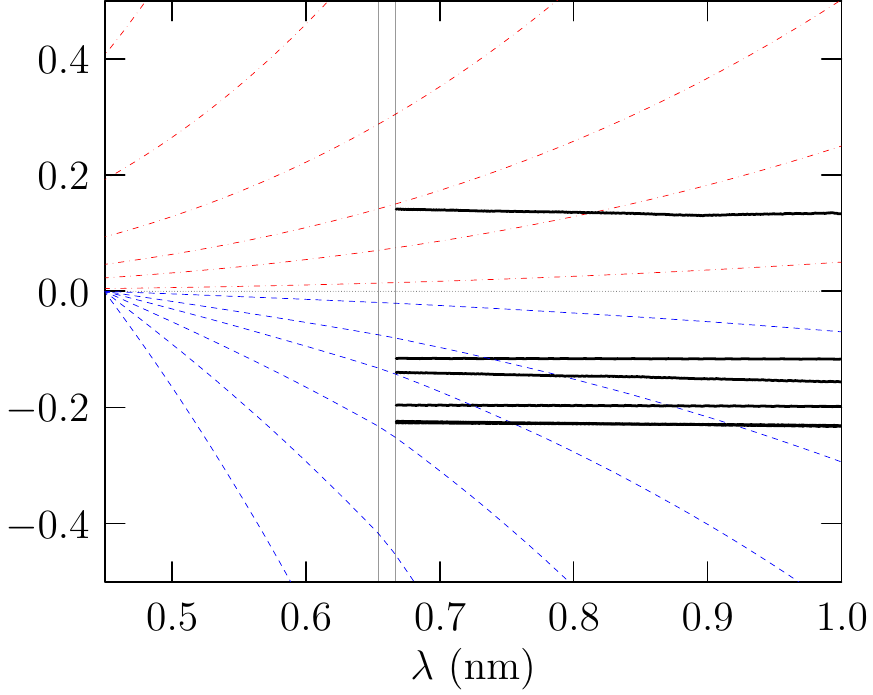}}
\caption{The red, blue, and black curves are the packing (arithmetic
average of $-\ln\left\langle\left\langle \chi
\right\rangle\right\rangle_0$ for the two ion types), chemical
(arithmetic average of $\ln \left\langle \chi \right\rangle$ for the 
two ion types), and net contributions 
(following Eq.~\eqref{eq:BIGQCT},
including the outer-shell contribution, for $\ln \gamma_\pm$),
respectively, for the primitive model results.\cite{Zhu:2011wn} The
various curves correspond to $c$ = \{0.01, 0.05, 0.1, 0.2, 0.4, 0.8\}
mol/dm$^3$ cases of Table~\ref{table:table2}.   
}
\label{fig:netconV1}
\end{figure}

\begin{figure}
\includegraphics[width=3.5in]{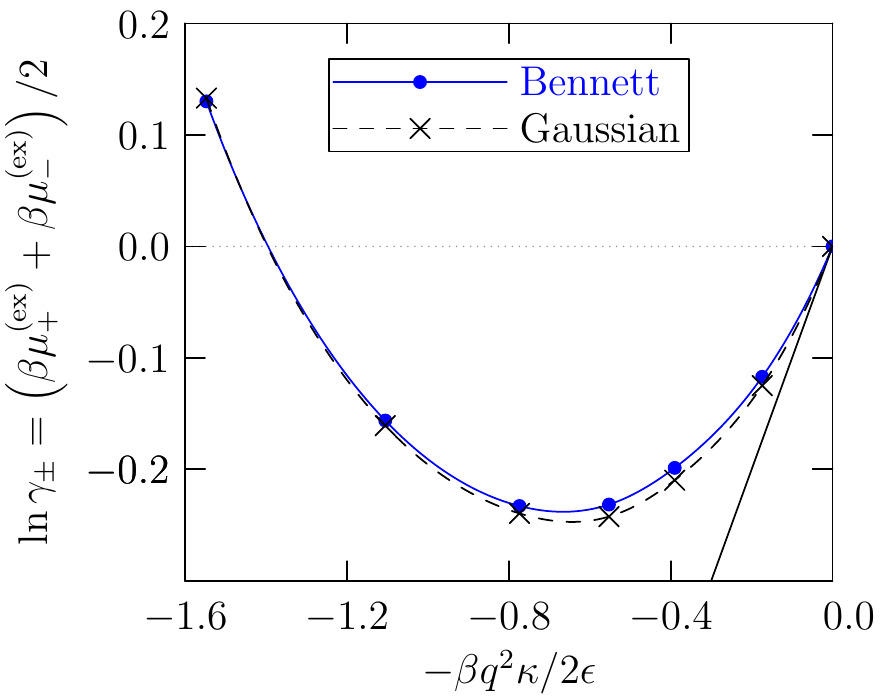}
\caption{QCT, as discussed in Sec.~\ref{sec:QCT}.
The solid black line is the Debye-H\"uckel limiting law.
}
\label{fig:meangammaV5lambda}
\end{figure}

Though the various QCT contributions depend  on the radius $\lambda$ of
the inner-shell (Fig.~\ref{fig:netconV1}), the net free energy varies
only slightly with increases of $\lambda > 0.7$~nm. The mean activity
coefficients evaluated by the Bennett method and the gaussian
approximation (Fig.~\ref{fig:meangammaV5lambda}) now accurately agree.
This suggests that both approaches are physically reliable with this
conditioning. In order that an MM pair-potential at long-range may be
plausibly exploited, the conditioning is essential to the broader
idea here .

For the concentrations and $\lambda$ values in Fig.~\ref{fig:netconV1},
the Poisson estimates of the packing and chemical
contributions\cite{Zhu:2011wn}
\begin{multline}
-\ln \left\langle\left\langle \chi \right\rangle\right\rangle_0 \equiv 
-\sum_\alpha \ln \left\langle\left\langle \chi_\alpha \right\rangle\right\rangle_0
\\
=
\left(\frac{1}{2}\right)\sum_{\alpha,\gamma} \frac{4\pi}{3}\lambda^3 c~,
\label{eq:p0Poisson}
\end{multline}
\begin{multline}
-\ln \left\langle
\chi \right\rangle  \equiv 
-\sum_\alpha \ln \left\langle \chi_\alpha \right\rangle
\\
=
\left(\frac{1}{2}\right)\sum_{\alpha,\gamma} 4\pi\int_{0}^\lambda  c g_{\alpha\gamma}\left(r\right)  r^2 \dif r~,
\label{eq:pPoisson}
\end{multline}
are useful.   The combination 
\begin{multline}
-\ln \left\lbrack \frac{\left\langle\left\langle \chi \right\rangle\right\rangle_0}{\left\langle
\chi \right\rangle}\right\rbrack = 
-  \left(\frac{1}{2}\right)\sum_{\alpha,\gamma} 4\pi \int_{0}^\lambda 
c\left\lbrack 
g_{\alpha\gamma}\left(r\right) -1
\right\rbrack r^2 \dif r~.
\end{multline}
is then interesting.  The utility of these results emphasize again
that the ion densities are not high, so simple results can be helpful.

\begin{figure}
\center{\includegraphics[width=1.0\linewidth]{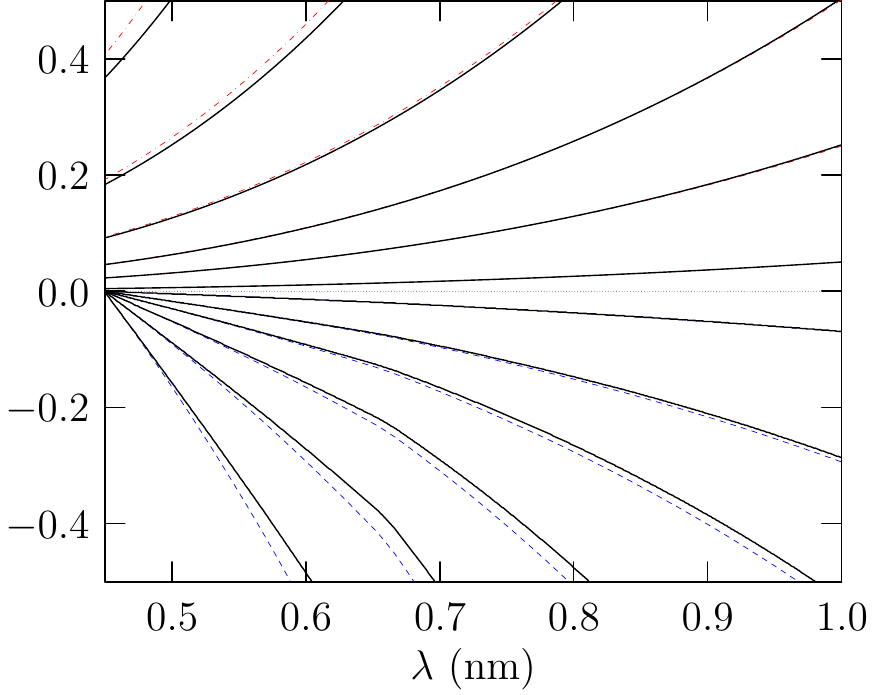}}
\caption{Upper black curves are the results fro Eq.~\eqref{eq:p0Poisson}.
Similarly, the red dot-dashed curves are the direct numerical 
results obtained by trail insertions.   The lower black 
are the results fro Eq.~\eqref{eq:pPoisson}.
Similarly, the blue dashed curves are the direct numerical 
results obtained by observations of the ions 
present in the simulations.
}
\label{fig:pcheck}
\end{figure}

\section{Conclusions}
This paper organizes several basic theoretical results ---
McMillan-Mayer theory, the potential distribution approach, and
quasi-chemical theory --- to apply high-resolution AIMD  to electrolyte
solutions. The conceptual target for these considerations is that the
last of the calculations depicted in Fig.~\ref{fig:QCT} can be done by
AIMD directly on the time and space scales typical of those demanding
methods. This theory then develops a mechanism for addressing effects
associated with longer spatial scales, involving also characteristically
longer time scales.   The theory treats composition fluctuations which
would be accessed by larger-scale calculations, and also longer-ranged
interactions and correlations that are of special interest for
electrolyte solutions.

The quasi-chemical organization, as an extension of van~der~Waals
pictures, breaks-up governing free energies into physically distinct
contributions: \emph{packing}, \emph{outer-shell}, and \emph{chemical}
contributions.   This paper adopted a primitive model suggested by
observed ion-pairing  in tetra-ethylammonium tetra-fluoroborate
dissolved in propylene carbonate, then studied specifically the
\emph{outer-shell} contributions that expresses electrolyte screening.
Gaussian statistical models are shown to be effective as physical models
for these \emph{outer-shell} contributions, and they are conclusive for
the free energies within the quasi-chemical formulation
(Figs.~\ref{fig:netconV1} and \ref{fig:meangammaV5lambda}).  In fact,
with this data set the gaussian physical approximation is more efficient
in providing an accurate mean activity coefficient than is the Bennett
direct evaluation of that free energy (Figs.~\ref{fig:meangammaV5} and
\ref{fig:meangammaV5lambda}).

\section{Acknowledgements}

This work was supported by the National Science Foundation under the NSF EPSCoR 
Cooperative Agreement No. EPS-1003897, with additional support from the Louisiana 
Board of Regents.

  \renewcommand{\theequation}{A-\arabic{equation}}
  \setcounter{equation}{0}  
  
\section*{Appendix A: Accessible derivation of the McMillan-Mayer Theory}\label{sec:appendix}
In the analysis of the MM theory, the formulae that are employed can be
intimidating\cite{mayer-mayer,munster,friedman} at several stages so a physically clear notation helps. We
consider a system composed of solvent (S) and solutes (A). The numbers
of these species will be indicated by $\vec{n}_{\mathrm{S}}$ and
$\vec{n}_{\mathrm{A}}$, the bold-face typography indicating that each of
these quantities can be multi-component, \emph{i.e.},
$\vec{n}_{\mathrm{A}} = \{ n_{\mathrm{A_1}},n_{\mathrm{A_2}}, \ldots\},
$ and similarly for solvent species.  The Helmholtz free energy
$A(T,V,\vec{n}_{\mathrm{S}},\vec{n}_{\mathrm{A}})$ then leads to the
canonical partition function
\begin{eqnarray}
\me^{-\beta A\left(T,V,\vec{n}_{\mathrm{S}},\vec{n}_{\mathrm{A}}\right)}  = 
\mathcal{Q}\left(\vec{n}_{\mathrm{S}},\vec{n}_{\mathrm{A}}\right)/\left(\vec{n}_{\mathrm{S}}!\vec{n}_{\mathrm{A}}!\right)
~.
\label{eq:helmholtzA}
\end{eqnarray}
$T$ (temperature) and $V$ (volume) have their usual meanings, and we
will suppress that notation on the right of Eq.~\eqref{eq:helmholtzA}.
To further fix the notation we recall\cite{BPP}
$\mathcal{Q}(n_\mathrm{A}=1)\equiv
Vq_\mathrm{A}^{\left(\mathrm{int}\right)}/\Lambda _\mathrm{A}{}^3$ is
the canonical ensemble partition function for a system comprising
exactly one molecule of type A in a volume $V$ with
$\Lambda_{\mathrm{A}}$ is the thermal deBroglie wavelength. The
factorial notation
\begin{eqnarray}
\vec{n}_{\mathrm{A}}! = n_{\mathrm{A_1}}! n_{\mathrm{A_2}}! \ldots
\end{eqnarray}
is common.\cite{BPP,mayer-mayer} When convenient, we will denote $\vec{n} =
\{\vec{n}_{\mathrm{S}},\vec{n}_{\mathrm{A}}\}$ so that 
\begin{eqnarray}
\me^{-\beta A\left(T,V,\vec{n}\right)}   =  
\mathcal{Q}\left(\vec{n}\right)/\vec{n}!
~.
\label{eq:helmholtzB}
\end{eqnarray}
The grand
canonical partition function will be central,
\begin{eqnarray}
\me^{\beta p V} = \sum_{\vec{n} \ge 0}\mathcal{Q}\left(\vec{n}\right)\left(\frac{\vec{z}{}^{\vec{n}}}{\vec{n}!}\right)~,
\label{eq:gcanonical}\end{eqnarray}
in these terms.  Here we adopt a correspondingly simplified notation for the activities\cite{BPP,mayer-mayer}
\begin{eqnarray}
\vec{z}{}^{\vec{n}} = \exp\left\lbrace\sum_\mathrm{X}\beta\mu_{\mathrm{X}}n_{\mathrm{X}} \right\rbrace
\end{eqnarray}
with $\mu_{\mathrm{X}}$ the chemical potential of species X.  We will
compare the pressure of the solution with the pressure of the
solvent-only system  at the same activity
$z_{\mathrm{S}}=\me^{\beta\mu_{\mathrm{S}}}$:
\begin{eqnarray}
\me^{\beta \left(p-\pi\right) V} = \sum_{\vec{n}_{\mathrm{S}} \ge 0}
\mathcal{Q}\left(\vec{n}_{\mathrm{S}},\vec{n}_{\mathrm{A}}=0\right)
\left(\frac{\vec{z}_{\mathrm{S}}{}^{\vec{n}_{\mathrm{S}}}}{\vec{n}_{\mathrm{S}}!}\right)~.
\end{eqnarray}
The pressure difference $\pi$ is the osmotic pressure.   The
probability for observing $\vec{n}_{\mathrm{S}}$ in the solvent-only
system is
\begin{multline}
P\left(\vec{n}_{\mathrm{S}};\vec{z}_{\mathrm{A}}=0\right) = 
\mathcal{Q}\left(\vec{n}_{\mathrm{S}},\vec{n}_{\mathrm{A}}=0\right)\left(\frac{\vec{z}_{\mathrm{S}}{}^{\vec{n}_{\mathrm{S}}}}{\vec{n}_{\mathrm{S}}!}\right)
\\
\times
\me^{-\beta \left(p-\pi\right) V}~.\end{multline}
With these notations we write
\begin{multline}
\me^{\beta \pi V}  = \\
\sum_{\vec{n}_{\mathrm{A}}\ge0}
\left(
\frac{
\vec{z}_{\mathrm{A}}{}^{\vec{n}_{\mathrm{A}}}
}{
\vec{n}_{\mathrm{A}}!
}
\right)
\sum_{\vec{n}_{\mathrm{S}}\ge0}\left\lbrace\frac{\mathcal{Q}\left(\vec{n}_{\mathrm{S}},\vec{n}_{\mathrm{A}}\right)}{\mathcal{Q}\left(\vec{n}_{\mathrm{S}},\vec{n}_{\mathrm{A}}=0\right)} \right\rbrace P\left(\vec{n}_{\mathrm{S}};\vec{z}_{\mathrm{A}}=0\right)
\\
 = \sum_{\vec{n}_{\mathrm{A}}\ge0}\mathcal{Z}\left(\vec{n}_{\mathrm{A}};\vec{z}_{\mathrm{S}}\right)
\left(
\frac{
\vec{z}_{\mathrm{A}}{}^{\vec{n}_{\mathrm{A}}}
}{
\vec{n}_{\mathrm{A}}!
}
\right)~.
\end{multline}
The important point is the structural similarity to
Eq.~\eqref{eq:gcanonical}.

Our task is to analyze the MM configurational integral 
\begin{multline}
\mathcal{Z}\left(\vec{n}_{\mathrm{A}};\vec{z}_{\mathrm{S}}\right)
= \\
\sum_{\vec{n}_{\mathrm{S}}\ge0}\left\lbrace\frac{\mathcal{Q}\left(\vec{n}_{\mathrm{S}},\vec{n}_{\mathrm{A}}\right)}{\mathcal{Q}\left(\vec{n}_{\mathrm{S}},\vec{n}_{\mathrm{A}}=0\right)} \right\rbrace 
P\left(\vec{n}_{\mathrm{S}};\vec{z}_{\mathrm{A}}=0\right)
\label{eq:MMPF}
\end{multline}
The displayed ratio of partition functions is distinctive.  For the case $\vec{n}_{\mathrm{A}}=1$,
for example, we  write
\begin{multline}
\sum_{\vec{n}_{\mathrm{S}}\ge0}\left\lbrace\frac{\mathcal{Q}\left(\vec{n}_{\mathrm{S}},\vec{n}_{\mathrm{A}}=1\right)}{\mathcal{Q}\left(\vec{n}_{\mathrm{S}},\vec{n}_{\mathrm{A}}=0\right)} \right\rbrace 
P\left(\vec{n}_{\mathrm{S}};\vec{z}_{\mathrm{A}}=0\right) \\ = 
\mathcal{Q}\left(\vec{n}_{\mathrm{S}}=0,\vec{n}_{\mathrm{A}}=1\right)
\left\langle\left\langle \me^{-\beta \Delta U_{\mathrm{A}}^{\left(1\right)}}\right\rangle\right\rangle_0~,
\label{eq:pdtexample}
\end{multline}
where the right-most factor is to be evaluated at infinite dilution of the solute,
$\vec{z}_{\mathrm{A}}=0$.
The potential distribution development establishes that right-side to be\cite{BPP}
\begin{multline}
\mathcal{Q}\left(\vec{n}_{\mathrm{S}}=0,\vec{n}_{\mathrm{A}}=1\right)\left\langle\left\langle \me^{-\beta \Delta U_{\mathrm{A}}^{\left(1\right)}}\right\rangle\right\rangle_0
\\
= \lim_{z_A \rightarrow 0} \left(\frac{n_\mathrm{A}}{z_A }\right)
= \lim_{z_A \rightarrow 0} \left(\frac{\rho_\mathrm{A}}{z_A }\right)V~.
\label{eq:pdtexamplenext}{}
\end{multline}

To write the general term for Eq.~\eqref{eq:MMPF}, we will use $${\vec{n}_{\mathrm{A}} \choose
\vec{m}_{\mathrm{A}}}$$ to denote the number of ways of selecting the
$\vec{m}_{\mathrm{A}}$ solute molecule set from the collection
$\vec{n}_{\mathrm{A}}$.   For example, if only one type of solute A is
considered, then
\begin{eqnarray}
{\vec{n}_{\mathrm{A}} \choose \vec{m}_{\mathrm{A}}} = 
\frac{n_\mathrm{A}!}{m_\mathrm{A}!\left(n_\mathrm{A}-m_\mathrm{A}\right)!} = 
\frac{n_\mathrm{A}^{\underline{m}_\mathrm{A}}}{m_\mathrm{A}!}~,
\end{eqnarray}
as usual, with the last equality using the
`$n_{\mathrm{A}}$-to-the-$m_{\mathrm{A}}$-falling'
notation.\cite{BPP,knuth}  

For more general but specified $\vec{m}_{\mathrm{A}}$, we rewrite Eq.~\eqref{eq:MMPF} 
\begin{multline}
\sum_{\vec{n}_{\mathrm{S}}\ge0}\left\lbrace\frac{\mathcal{Q}\left(\vec{n}_{\mathrm{S}},\vec{m}_{\mathrm{A}}\right)
}{
\mathcal{Q}\left(\vec{n}_{\mathrm{S}},\vec{m}_{\mathrm{A}}=0\right)} \right\rbrace 
P\left(\vec{n}_{\mathrm{S}};\vec{z}_{\mathrm{A}}=0\right) \\
= \mathcal{Q}\left(\vec{n}_{\mathrm{S}}=0,\vec{m}_{\mathrm{A}}\right)\left\langle\left\langle \me^{-\beta \Delta U^{\left(\vec{m}_{\mathrm{A}}\right)}}\right\rangle\right\rangle_0
\label{eq:MMPFnext}
\end{multline}
and again, after having set $\vec{m}_{\mathrm{A}}$, this is to be
evaluated at infinite dilution.   Here the binding energy
\begin{multline}
\Delta U^{\left(\vec{m}_{\mathrm{A}}\right)} = U\left(\vec{n}_{\mathrm{S}},\vec{m}_{\mathrm{A}}\right) \\
 - U\left(\vec{n}_{\mathrm{S}},\vec{m}_{\mathrm{A}}=0\right) 
 -	U\left(\vec{n}_{\mathrm{S}}=0,\vec{m}_{\mathrm{A}}\right)~,
\end{multline}
is associated with the collection of $\vec{m}_{\mathrm{A}}$ solute molecules.  Following the potential
distribution theory further \cite{BPP}
\begin{multline}
\mathcal{Q}\left(\vec{n}_{\mathrm{S}}=0,\vec{m}_{\mathrm{A}}\right)
\left\langle\left\langle \me^{-\beta \Delta U^{\left(\vec{m}_{\mathrm{A}}\right)}}\right\rangle\right\rangle_0 \\
= 
\left\langle {\vec{n}_{\mathrm{A}} \choose \vec{m}_{\mathrm{A}}}\right\rangle
\frac{\vec{m}_{\mathrm{A}}!}{\vec{z}_{\mathrm{A}}{}^{\vec{m}_{\mathrm{A}}}}~,
\label{eq:interesting}
\end{multline}   
Finally, 
\begin{multline}
\left\langle {\vec{n}_{\mathrm{A}} \choose \vec{m}_{\mathrm{A}}}\right\rangle \vec{m}_{\mathrm{A}}!\\
= \vec{\rho}_{\mathrm{A}}{}^{\vec{m}_{\mathrm{A}}}\int_V \dif 1_\mathrm{A}
\ldots \int_V \dif m_\mathrm{A} g^{{\left(\vec{m}_{\mathrm{A}}\right)}}\left( 1_\mathrm{A} \ldots
m_\mathrm{A} \right)~,\end{multline}
with $g^{{\left(\vec{m}_{\mathrm{A}}\right)}}\left( 1_\mathrm{A} \ldots
m_\mathrm{A} \right)$ denoting the usual $\vec{m}_{\mathrm{A}}$ joint
distribution function. Here we denote solute configurational coordinates
as $\left( 1_\mathrm{A}, \ldots m_\mathrm{A} \right)$, and the necessary
integrations by $\int_V \dif 1_\mathrm{A} \ldots \int_V \dif
m_\mathrm{A}$. This produces the factor of $V$ in
Eq.~\eqref{eq:pdtexamplenext}.  Since we wish to simplify
Eq.~\eqref{eq:MMPFnext}, with $\vec{z}_{\mathrm{A}}=0$, we use
Eq.~\eqref{eq:interesting} to write
\begin{multline}
\mathcal{Z}\left(\vec{n}_{\mathrm{A}};\vec{z}_{\mathrm{S}}\right)
 = 
\left\lbrack \lim_{\vec{z}_{\mathrm{A}}\rightarrow 0}\left(\frac{\vec{\rho}_{\mathrm{A}}}{\vec{z}_{\mathrm{A}}}\right)\right\rbrack^{\vec{n}_{\mathrm{A}}} 
\\
\int_V \dif 1_\mathrm{A}
\ldots \int_V \dif n_\mathrm{A} g^{{\left(\vec{n}_{\mathrm{A}}\right)}}\left( 1_\mathrm{A} \ldots
n_\mathrm{A};\vec{z}_{\mathrm{A}}=0 \right)
\end{multline}
The prefactor, to be evaluated at infinite dilution,
is given by 
\begin{eqnarray}
\frac{\rho_\mathrm{A}}{z_\mathrm{A}} = \frac{q_\mathrm{A}^{\mathrm{int}}}{\Lambda_\mathrm{A}^3}
\left\langle\left\langle
\me^{-\beta \Delta U^{(1)}_\mathrm{A}}\right\rangle
\right\rangle_0
\label{eq:rawPDT}
\end{eqnarray}
in the potential distribution theorem formulation.\cite{BPP}

With this suggestive form we can be more specific about the 
canonical configurational integrals that started our discussion, 
specifically
\begin{multline}
\mathcal{Q}\left(\vec{n}_{\mathrm{A}}\right) = \lim_{\vec{z}_\mathrm{S} \rightarrow 0} 
\mathcal{Z}\left(\vec{n}_{\mathrm{A}};\vec{z}_\mathrm{S}\right) \\
= \lim_{\vec{z}_\mathrm{S} \rightarrow 0}
\left\lbrack
\lim_{\vec{z}_{\mathrm{A}}\rightarrow 0}\left(\frac{\rho_\mathrm{A}}{z_\mathrm{A}}\right)\right\rbrack^{\vec{n}_{\mathrm{A}}} \\
\times 
\int_V \dif 1_\mathrm{A}
\ldots \int_V \dif n_\mathrm{A}\me^{ - \beta W\left( 1_\mathrm{A} \ldots
n_\mathrm{A} \right)}~.
\label{eq:specific}
\end{multline}
The multipliers appearing on the middle line supply features of
the kinetic energy portion of the partition function, specific to the 
implementation for the particular case.  For notational simplicity we
will drop the specific identification of the solvent activity in the
formulae elsewhere.

These formulae, particularly Eq.~\eqref{eq:specific}, are collected 
in the summary statement of MM  theory in Sec.~\ref{sec:MM}, and
particularly with Eq.~\eqref{eq:MMpfFINAL}.

  \renewcommand{\theequation}{B-\arabic{equation}}
  \setcounter{equation}{0}  
  
\section*{Appendix B: Potential Distribution Theory}\label{sec:appendixB}
With the MM background, we  evaluate the average number of solute A molecules as
\begin{eqnarray}
\left \langle n_{\mathrm{A}} \right \rangle = 
\me^{-\beta \pi V}
\sum_{\vec{n}_{\mathrm{A}}\geq 0}^{ } n_{\mathrm{A}} 
\mathcal{Z}\left ( \vec{n}_{\mathrm{A}} ;\vec{z}_{\mathrm{S}}\right )
\left(\frac{\vec{z}_\mathrm{A}{}^{\vec{n}_\mathrm{A}}}{\vec{n}_\mathrm{A}!}\right)~.
\end{eqnarray}
Since the summand factor $n_{\mathrm{A}}$ annuls the
$n_{\mathrm{A}} = 0$ term, this result presents
an explicit leading factor of $z_{\mathrm{A}}$. Determination
of $z_{\mathrm{A}}$  establishes the
thermodynamic property $\mu_{\mathrm{A}}$. Therefore, we
rewrite this equation by bringing forward the explicit extra factor of
$z_{\mathrm{A}}$ as
\begin{widetext}
\begin{eqnarray}
\left \langle n_{\mathrm{A}} \right \rangle = e^{-\beta \pi V}
\mathcal{Z}\left ( \vec{n}_{\mathrm{A}} = 1; \vec{z}_{\mathrm{S}} \right) z_{\mathrm{A}} 
\sum_{\vec{n}_{\mathrm{A}}\geq 0}^{ }
\left ( 
\frac{\mathcal{Z}\left ( \vec{n}_{\mathrm{A}} + 1; \vec{z}_{\mathrm{S}} \right )
}{
\mathcal{Z}\left ( \vec{n}_{\mathrm{A}} = 1; \vec{z}_{\mathrm{S}} \right)\mathcal{Z}\left ( \vec{n}_{\mathrm{A}}; \vec{z}_{\mathrm{S}} \right )
} 
\right ) 
\mathcal{Z}\left ( \vec{n}_{\mathrm{A}}; \vec{z}_{\mathrm{S}} \right )
\left(\frac{\vec{z}_\mathrm{A}{}^{\vec{n}_\mathrm{A}}}{\vec{n}_\mathrm{A}!}\right)~.
\end{eqnarray}
\end{widetext}
or 
\begin{eqnarray}
\left \langle n_{\mathrm{A}} \right \rangle = \mathcal{Z}\left ( \vec{n}_{\mathrm{A}} = 1; \vec{z}_{\mathrm{S}} \right ) z_{\mathrm{A}}
\left\langle\left\langle
\me^{-\beta \Delta W^{(1)}_\mathrm{A}}\right\rangle
\right\rangle_0~.
\label{eq:MMmeann}
\end{eqnarray}
Here 
\begin{eqnarray}
\Delta W^{(1)}_\mathrm{A} = W\left(\vec{n}_A+1\right) - W\left(\vec{n}_A\right) - W\left(1\right)~,
\end{eqnarray}
is the binding energy of a distinguished solute (A) molecule in the MM
system, and the quantity
\begin{eqnarray}
\mathcal{Z}\left ( \vec{n}_{\mathrm{A}} = 1; \vec{z}_{\mathrm{S}} \right ) = \frac{V q_\mathrm{A}^{\left(\mathrm{int}\right)}}{\Lambda_\mathrm{A}{}^3}
\left\langle\left\langle
\me^{-\beta \Delta U^{(1)}_\mathrm{A}}\right\rangle
\right\rangle_0~
\end{eqnarray}
involves interactions of one A molecule and the solvent; it is proportional to
the system volume.

  \renewcommand{\theequation}{C-\arabic{equation}}
  \setcounter{equation}{0}  
  
\section*{Appendix C: QCT breakup in the grand canonical ensemble}\label{sec:appendixC}
Here we discuss twists associated with the
consideration of PDT developments when $n_\mathrm{A}$ fluctuates.
We begin with the observation from Eq.~\eqref{eq:MMmeann} that
\begin{eqnarray}
\left\langle\left\langle
\me^{-\beta \Delta W^{(1)}_\mathrm{A}}\right\rangle
\right\rangle_0 \propto \left \langle n_{\mathrm{A}} \right \rangle~.
\label{eq:MMmeann2}
\end{eqnarray}
Then considering the ratio
\begin{eqnarray}
 \frac{
\left\langle\left\langle
\me^{- \beta \Delta W^{(1)}_\mathrm{A}} F
\right\rangle\right\rangle_0
}{
\left\langle\left\langle
\me^{- \beta \Delta W^{(1)}_\mathrm{A}}
\right\rangle\right\rangle_0
} = \frac{
\left\langle F n_\mathrm{A} \right\rangle
}{
\left\langle n_\mathrm{A} \right\rangle
}~,
\end{eqnarray}
yields a particularly transparent result. Choosing $ F=\me^{\beta \Delta
W^{(1)}_\mathrm{A}}\chi_{\mathrm{A}}$, we obtain an analogue of Eq~\eqref{eq:conditioning}:
\begin{eqnarray}
\frac{\left\langle\me^{ \beta \Delta W^{(1)}_\mathrm{A}}\chi_{\mathrm{A}} n_\mathrm{A} \right\rangle
}{\left\langle n_\mathrm{A} \right\rangle
}= \frac{
\left\langle\left\langle
\chi_{\mathrm{A}}
\right\rangle\right\rangle_0
}{
\left\langle\left\langle
\me^{- \beta \Delta W^{(1)}_\mathrm{A}}
\right\rangle\right\rangle_0
}~.
\label{eq:n-average}
\end{eqnarray}
If the averages are canonical then this is just
Eq~\eqref{eq:conditioning} again, but Eq.~\eqref{eq:n-average} remains
true if $n_\mathrm{A}$ fluctuates. 

We expect that 
\begin{multline}
\frac{\left\langle\me^{ \beta \Delta W^{(1)}_\mathrm{A}}\chi_{\mathrm{A}} n_\mathrm{A} \right\rangle
}{\left\langle n_\mathrm{A} \right\rangle
} \sim  \left\langle\me^{ \beta \Delta W^{(1)}_\mathrm{A}}\chi_{\mathrm{A}} \right\rangle \\
 + O(\left\langle n_\mathrm{A} \right\rangle{}^{-1})~,
\label{eq:asymptotic}
\end{multline}
so in the thermodynamic limit that average matches the simpler canonical
expression. The physical reason for this expectation is that we can
write $ n_\mathrm{A} = \left\langle n_\mathrm{A} \right\rangle + \delta
n_\mathrm{A}$ in the numerator.   Then the correlation of $\delta
n_\mathrm{A}$ with the intensive characteristic of that numerator
average should yield an intensive result.

Accepting this argument for the moment and retaining only 
the dominant contribution in Eq.~\eqref{eq:asymptotic}, we
recover the results of Sec.~\ref{sec:QCT} --- and 
specifically the important result Eq.~\eqref{eq:BIGQCT} ---  
but consistently with the grand canonical ensemble derivation 
of the earlier sections.

To make that physical view specific, we introduce the additional
notation $$\left\langle \me^{\beta \Delta W^{(1)}_\mathrm{A}}\vert
n_\mathrm{A}\right\rangle$$ for the canonical ensemble average that
specifies $n_\mathrm{A}$. For anticipated $\delta n_\mathrm{A}$, we use
\begin{multline}
\left\langle \me^{\beta \Delta W^{(1)}}\chi_{\mathrm{A}}\vert n_\mathrm{A}\right\rangle
\approx 
\left\langle \me^{\beta \Delta W^{(1)}_\mathrm{A}}\chi_{\mathrm{A}}\vert \left\langle n_\mathrm{A}\right\rangle{}\right\rangle \\
 + \delta n_\mathrm{A}\left( \frac{\partial \left\langle \me^{\beta \Delta W^{(1)}_\mathrm{A}}\chi_{\mathrm{A}}\vert \left\langle n_\mathrm{A}\right\rangle{}\right\rangle}{\partial\left\langle n_\mathrm{A}\right\rangle}\right)~.
\end{multline}
Used in the left-side of Eq.~\eqref{eq:asymptotic}, and then averaging
with respect to $n_{\mathrm{A}}$ occupancies, this yields
\begin{multline}
\frac{\left\langle\me^{ \beta \Delta W^{(1)}_\mathrm{A}}\chi_{\mathrm{A}} n_\mathrm{A} \right\rangle
}{\left\langle n_\mathrm{A} \right\rangle
} \approx
\left\langle \me^{\beta \Delta W^{(1)}_\mathrm{A}}\chi_{\mathrm{A}}\vert \left\langle n_\mathrm{A}\right\rangle{}\right\rangle \\
 + \frac{\left\langle\delta n_\mathrm{A}{}^2\right\rangle
 }{
 \left\langle n_\mathrm{A} \right\rangle}
 \left( \frac{\partial \left\langle \me^{\beta \Delta W^{(1)}_\mathrm{A}}\chi_{\mathrm{A}}\vert \left\langle n_\mathrm{A}\right\rangle{}\right\rangle}{\partial\left\langle n_\mathrm{A}\right\rangle}\right)~,
\label{eq:finally}
\end{multline}
the expected result. Since $$\left\langle\delta
n_\mathrm{A}{}^2\right\rangle = \left(\frac{\partial \left\langle
n_\mathrm{A} \right\rangle}{\partial \beta
\mu_\mathrm{A}}\right)_{T,V,\mu_{\mathrm{S}}}~,$$ the correction indeed vanishes in the
thermodynamic limit.

  \renewcommand{\theequation}{D-\arabic{equation}}
  \setcounter{equation}{0}  


\pagebreak

%

\end{document}